\documentstyle[12pt]{article}
\begin{document}

\title{Quantum computation with trapped ions in strongly detuned optical cavity.}
\author{XuBo Zou, K. Pahlke and W. Mathis  \\
\\Institute TET, University of Hannover,\\
Appelstr. 9A, 30167 Hannover, Germany }
\date{}

\maketitle

\begin{abstract}
{\normalsize We propose a scheme to implement quantum phase gate
for two $\Lambda$ ions trapped in optical cavity. It is shown that
quantum phase gate can be implemented by applying a laser
addressing to a single ions in strongly detuned optical cavity. We
further demonstrate that geometric quantum phase gate can be
implemented by introducing a auxiliary ground state.  }

PACS number(s): 03.67.-a, 03.67.Hk.
\end{abstract}
In recent years, it is realized that quantum computer can provide
more powerful computational ability than a classical
one\cite{shor,gro}. This discovery motivated intensive research
into apparatus which can be used to perform quantum logic
operation on single or multiple qubits. To bulid a real quantum
computer, we need to find quantum system, which have little
decoherence and easy way for coherent operation. Several possible
physical implementation have been suggested, including the Cavity
QED\cite{tc}, trapped ion system\cite{cd} and NMR system\cite{cg}
and solid state system using nuclear spins\cite{ka}, quantum
dots\cite{loss} and Josephson Junction\cite{mak}. The ion trap
quantum computation, first proposed by Cirac and Zoller\cite{cz},
and demonstrated experimentally shortly afterward\cite{cm}, is a
promising candidate for realizing a small scale of quantum
computation. They offer the possibility in principle to store
string of ions, cool them almost to zero temperature and address
them individually with laser field. The original idea of Cirac and
Zoller uses the collective spatial vibrations for communication
between ions, and it requires that the system is restricted to the
joint motional ground state of the ions. For two ions, this has
recently been accomplished\cite{bek}]. Any heating (i.e.
excitation by external fields) will diminish the accuracy of a
logic gate, thus leading to unreliable performance of the quantum
computer as a whole, and may be making the implement of even
simple algorithms impossible. Eliminating all of the possible
cause of heating is a very demanding task. Several
schemes\cite{pcz,sm,sjj,gjm,james} have been proposed to address
the problem. Recently approaches have been proposed which employs
ions trapped in an optical cavity. It has been shown that cavity
decay can be used to create entanglement implement quantum
computation between ions in cavity\cite{bbb}. More recently, a
scheme is proposed to implement quantum information processing
with ions trapped in strongly detuned optical cavity\cite{eee}.
All these scheme require that laser fields addressing different
ions was will matched in amplitude and phase. In this paper, we
propose a scheme to implement quantum logic gate (quantum phase
gate) for ions trapped in strongly detuned optical cavity by using
only
laser field addressing a single ion.\\
We first consider two Lambda-ions trapped in a cavity. It is
assumed that ions have the lower state $|0>$ and $|1>$, which can
be represented by different hyperfine level or zeeman levels, and
an excited state $|3>$ coupled individually to each ground state
by laser radiation with different polarization or frequencies. The
atom interact with each other via the common cavity radiation
field. Here we assume the ions are separated by at least one
optical wavelength in order to manipulate each ions by using laser
field individually and the optical cavity couple the states $|0>$
and $|3>$ of each ions. A laser is applied to derive ion1 between
state states $|1>$ and $|3>$ with amplitude $\Omega$. The
Hamiltonian describing the combined cavity-ion system can be
written as
$$
H=\omega_0(\sigma_{00}^1+\sigma_{00}^2)+\omega_3(\sigma_{33}^1+\sigma_{33}^2)+\omega_ca^{\dagger}a
$$
$$
+ga^{\dagger}(\sigma_{03}^1+\sigma_{03}^2)+ga(\sigma_{30}^1+\sigma_{30}^2)
+\Omega(\sigma_{31}^1+\sigma_{13}^1)
$$
where $\sigma_{ij}^k=|i>_k{_k<}j|$. $g$ are cavity-ion coupling
constant for two ions which is assumed to be equal. $a$ and
$a^{\dagger}$ are the annihilation and creation operator for the
cavity photons. The cavity field suffers irreversible losses on
account of leakage out of the cavity described by the Liouvillan
$$
\Lambda_f\rho=k(2a\rho{a^{\dagger}}-a^{\dagger}a\rho-\rho{a^{\dagger}a})
$$
where $2k$ is the rate of the loss of the photon. The dissipation
due to atomic spontaneous emission is given by
$$
\Lambda_a\rho=\tau(2\sum_{i=1,2}\sigma_{03}^i\rho{\sigma_{30}^i}-\sigma_{30}^i\sigma_{03}^i\rho-\rho{\sigma_{30}^i\sigma_{03}^i})
$$
$$
+\tau(2\sum_{i=1,2}\sigma_{13}^i\rho{\sigma_{31}^i}-\sigma_{31}^i\sigma_{13}^i\rho-\rho{\sigma_{31}^i\sigma_{13}^i})
$$
where $\tau$ is the rate of the atomic spontaneous emission. The
evolution of the system is given by
$$
\frac{d\rho}{dt}=-i[H,\rho]+\Lambda_f\rho+\Lambda_a\rho
$$
We consider the case of large detuning
$\Delta=\omega_c-\omega_3+\omega_0$, i.e. $\Delta>>g$. In this
case, there is no energy exchange between atom system and the
cavity and we obtain effective evolution of the system for
initially vacuum cavity field (in the interaction
picture)\cite{agar}
$$
\frac{d\rho}{dt}=-i[\Omega(\sigma_{31}^1+\sigma_{13}^1),\rho]+\frac{g^2}{k^2+\Delta^2}\{-i\Delta[(\sigma_{33}^1+\sigma_{33}^2+\sigma_{30}^1\sigma_{03}^2+\sigma_{03}^1\sigma_{30}^2)),\rho]
$$
$$
+k[2(\sum_{i=1,2}\sigma_{03}^i)\rho{(\sum_{i=1,2}\sigma_{30}^i)}-(\sum_{i=1,2}\sigma_{30}^i)(\sum_{i=1,2}\sigma_{03}^i)\rho
$$
$$
-\rho{(\sum_{i=1,2}\sigma_{30}^i)(\sum_{i=1,2}\sigma_{03}^i)})\}
+\Lambda_a\rho
$$
If the condition $\Delta>>k$ and $\frac{g^2}{\Delta}>>\tau$ are
satisfied, then the contribution due to damping in Eq.() is
negligibly small and it reduce to
$$
\frac{d\rho}{dt}=\frac{-ig^2\Delta}{k^2+\Delta^2}\Delta[(\sigma_{33}^1+\sigma_{33}^2+\sigma_{30}^1\sigma_{03}^2+\sigma_{03}^1\sigma_{30}^2)),\rho]
-i[\Omega(\sigma_{31}^1+\sigma_{13}^1),\rho]
$$
which shows that in a cavity, highly detuned from the atomic
transition frequency, the evolution of the system is given by the
effective Hamiltonian
$$
H_{eff}=\frac{g^2}{\Delta}[(\sigma_{33}^1+\sigma_{33}^2+\sigma_{30}^1\sigma_{03}^2+\sigma_{03}^1\sigma_{30}^2)]
+\Omega(\sigma_{31}^1+\sigma_{13}^1)
$$
In the bases $\{|00>, |01>, |10>, |11>, |31>, |13>, |33>,
|\phi_+>, |\phi_->\}$, where
$|\phi_{\pm}>=\frac{1}{\sqrt{2}}(|30>\pm|03>)$, the effective
Hamiltonian can be rewritten in the form
$$
H_{eff}=\frac{g^2}{\Delta}(2|33><33|+2|\phi_+><\phi_+|+|31><31|+|13><13|)
$$
$$
+\Omega(|33><13|+|31><11|+\frac{1}{\sqrt{2}}|\phi_+><10|+\frac{1}{\sqrt{2}}|\phi_-><10|)
$$
If the condition $\frac{g^2}{\Delta},
>>\Omega$ is satisfied, we can obtain the
effective Hamiltonian
$$
H=\frac{\Omega}{\sqrt{2}}(|\psi_-><10|+|10><\psi_-|)
$$
which can be used to implement quantum phase gate at the time
$t=\frac{\pi\sqrt{2}}{\Omega}$
$$
|00>\rightarrow|00>;|01>\rightarrow|01>
$$
$$
|10>\rightarrow-|00>;|11>\rightarrow|11>
$$
This quantum phase gate can be used to implement universal quantum
computation by combining single qubit quantum operation.\\
We further show that geometric conditional phase gate\cite{geo}
can be implemented by introducing a auxiliary ground state. We
assume the vacuum field is tuned along the $0-3$ transtion with
equal atom-cavity coupling $g$ for two ions. Consider a laser
tuned between an auxiliary ground state $|2>$ and $|3>$ of ion 1
with amplitude $\Omega_2$ and a second laser tuned between $|1>$
and $|3>$ of ion 1 with amplitude $\Omega_1$. If the condition
$\Delta>>g$ and $\frac{g^2}{\Delta},
>>|\Omega_1|, |\Omega_2|$ is satisfied, we can
obtain thee effective Hamiltonian
$$
H=\frac{1}{\sqrt{2}}(\Omega_2|\psi_1><20|+\Omega_1|\psi_1><10|+h.c.)
$$
To get geometric operation, we set
$|\Omega_1|/|\Omega_2|=\tan(\theta/2)$ and with phase difference
$\varphi=\varphi_1-\varphi_2$ with the control parameters $\theta,
\varphi$ undergoing a cyclic adiabatic evolution on the $(\theta,
\varphi )$ plane described by a loop $C$ from $\theta=0$. During
the evolution, the state $|00>, |01>, |11>$ are decoupled from
Hamiltonian(), while the state $|10>$ adiabatically follows as
$\cos\frac{\theta}{2}|10>+\sin\frac{\theta}{2}e^{i\varphi}|20>$,
which acquire a geometric phase $e^{i\varphi_{Berry}}|10>$, where
$\varphi_{Berry}=\int\sin\theta{d\theta{d\varphi}}$ and the
integration runs over the surface the loop $C$ enclose.\\
In summary, we have presented scheme to implement the two qubit
conditional phase gate for ion trapped in a far detuned optical
cavity. Together with single qubit rotation they consist a
universal set of gates. The cavity decay is effective suppressed
by the large detuning between the cavity field and ion. We also
condition where atomic sponaneous emission is negligible. In the
Ref\cite{eee}, two lasers are applied to derive two ions, which
require suitable amplitude and phases. Here we show it is enough
to implement quantum phase gate by using a laser field to derive
one of the ions.  We further demonstrate that geometric quantum
phase gate can be implemented by introducing a auxiliary ground
state.

\end{document}